\documentclass[journal=nalefd, manuscript=letter, layout=twocolumn]{achemso}
\setkeys{acs}{articletitle = true}
\usepackage{latexsym}
\usepackage{graphicx}
\usepackage{siunitx}

\title{Superconducting proximity effect in InAsSb surface quantum wells with in-situ Al contacts}

\author{William~Mayer}
\author{William~F.~Schiela}
\author{Joseph~Yuan}
\author{Mehdi~Hatefipour}
\affiliation{Center for Quantum Phenomena, Department of Physics, New York University, NY 10003, USA}

\author{Wendy~L.~Sarney}
\author{Stefan~P.~Svensson}
\author{Asher C.~Leff}
\affiliation{US Army Combat Capabilities Command, Army Research Laboratory, Adelphi, MD 20783, USA}

\author{Tiago Campos}
\affiliation{Department of Physics, University at Buffalo, State University of New York, Buffalo, New York 14260, USA}

\author{Kaushini~S.~Wickramasinghe}
\affiliation{Center for Quantum Phenomena, Department of Physics, New York University, NY 10003, USA}
\author{Matthieu~C.~Dartiailh}
\affiliation{Center for Quantum Phenomena, Department of Physics, New York University, NY 10003, USA}

\author{Igor \v{Z}uti\'c}
\affiliation{Department of Physics, University at Buffalo, State University of New York, Buffalo, New York 14260, USA}
\author{Javad~Shabani}
\email{jshabani@nyu.edu}
\affiliation{Center for Quantum Phenomena, Department of Physics, New York University, NY 10003, USA}

\begin{document}
\begin{abstract}

\textbf{We demonstrate a robust superconducting proximity effect in InAs$_{0.5}$Sb$_{0.5}$ quantum wells grown with epitaxial Al contacts, which has important implications for mesoscopic and topological superconductivity. Unlike more commonly studied InAs and InSb semiconductors, bulk InAs$_{0.5}$Sb$_{0.5}$ supports stronger spin-orbit coupling and a larger $g$-factor. However, these potentially desirable properties have not been previously measured in epitaxial heterostructures with superconductors, which could serve as a platform for fault-tolerant topological quantum computing. Through structural and transport characterization we observe high-quality interfaces and strong spin-orbit coupling. We fabricate Josephson junctions based on InAs$_{0.5}$Sb$_{0.5}$ quantum wells and observe a strong proximity effect. With a contact separation of \SI{500}{nm}, these junctions exhibit products $I_{c}R_{N} = \SI{270}{\micro V}$ and $I_{ex}R_{N} = \SI{230}{\micro V}$ of normal resistance $R_N$, critical current $I_c$, and excess current $I_{ex}$. Both of these quantities demonstrate a robust and long-range proximity effect with highly-transparent contacts.}

\end{abstract}

A given material can be transformed through proximity effects whereby it acquires correlations from its neighbors, for example, becoming superconducting or magnetic. Such proximity effects not only complement the conventional methods of designing materials by doping or functionalization, but can also overcome their various limitations and enable novel states of matter \cite{IgorReview}. A striking example of this approach is semiconductors with strong spin-orbit coupling (SOC) and large $g$-factor, in proximity to conventional superconductors. Such structures are predicted to support topological superconductivity with exotic quasi-particle excitations including Majorana bound states (MBS), which hold promise for fault-tolerant quantum computing \cite{,Alicea_2012,Sestoft18}. Through braiding (exchange) of MBS it is possible to reveal their peculiar non-Abelian statistics and implement fault-tolerant quantum gates \cite{Aasen2016}.

Most efforts to realize MBS have been focused on one-dimensional (1D) systems, typically relying on proximitized InAs and InSb nanowires in an applied magnetic field. However, their geometry has inherent difficulties to implement braiding and imposes strong constraints on material parameters to achieve topological superconductivity, usually inferred from observation of a quantized zero-bias conductance peak. Instead, to overcome these limitations there is a growing interest in 2D platforms of proximitized semiconductors, which would also support topological superconductivity. These advantages have recently been demonstrated in planar Josephson junctions \cite{Mayer19TopoTransition,Fornieri2019,Ren2019} where the phase transition between trivial and topological superconductivity can be tuned using gate voltages and the superconducting phase. This allows for more complicated networks that could support fusion, braiding, and large-scale Majorana manipulation.

The motivation to study InAs$_{0.5}$Sb$_{0.5}$ goes beyond proximity effects. In the past it was recognized as an important material for infrared applications \cite{detector}, but there remains limited data available on quantum transport \cite{Sestoft18}. More recently, the discovery of ultrafast lasers with spin-polarized carriers \cite{SpinLaserNature} calls for semiconductors with very short spin-relaxation times which is also expected from InAs$_{0.5}$Sb$_{0.5}$ considering its strong SOC.

To address this situation we use molecular beam epitaxy to fabricate high-quality 2D junctions of aluminum and InAs$_{0.5}$Sb$_{0.5}$. Our sample characterization and transport measurements in the normal and superconducting state are complemented by numerical analysis focusing on the role of the SOC and the quantum confinement which directly influence the resulting $g$-factors. While most of the implications of our work are given in the context of a possible platform for realizing superconducting junctions that could support Majorana bounds states, we expect that the presented results could also motivate further work in elucidating the normal-state properties of InAs$_{0.5}$Sb$_{0.5}$ and its applications. 

Our experiments on InAs$_{0.5}$Sb$_{0.5}$-based two-dimensional electron gas (2DEG) are accompanied by numerical studies of its electronic structure, Rashba SOC and $g$-factor. From previous work, it is reported that InAs$_{0.5}$Sb$_{0.5}$ can exhibit significantly larger spin-splitting \cite{Sestoft18}, compared to InAs or InSb in which transport properties have been extensively explored. The bulk $g$-factor of InAs$_{0.5}$Sb$_{0.5}$ is expected to reach up to -120 and exhibit SOC almost an order of magnitude stronger than InAs \cite{Winkler2016}. We find that the $g$-factor is suppressed in narrow quantum wells, while the linear term in spin-orbit coupling decreases as quantum well width is increased. 

\begin{figure*}
\begin{center}
\includegraphics[width=1\textwidth]{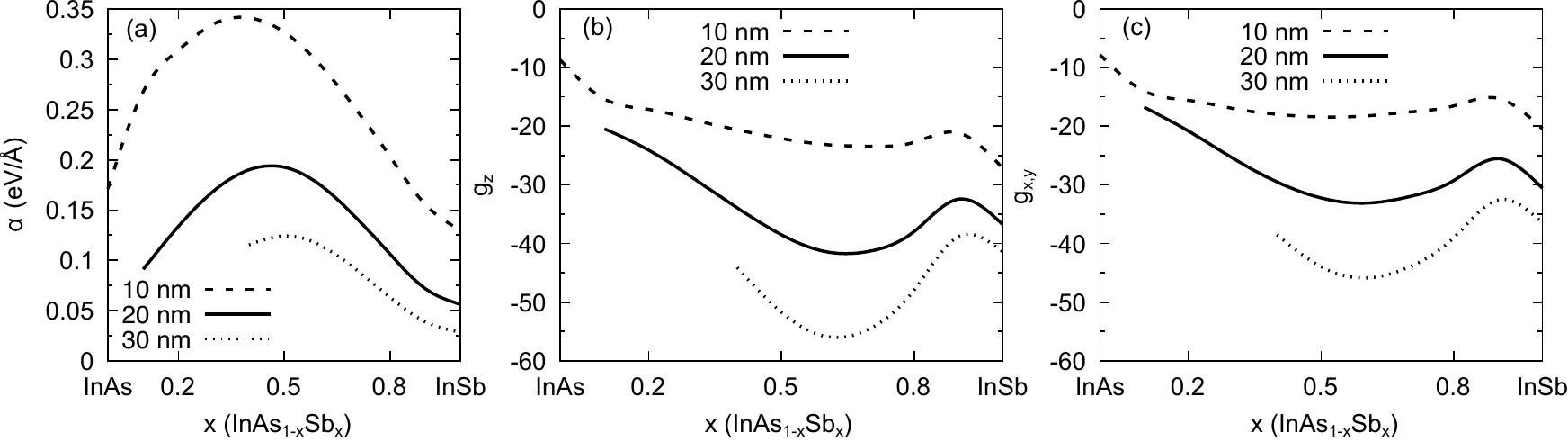}
\caption{Numerical calculations using the standard 8-band $\vec{k}\cdot\vec{p}$ method. Legends indicate quantum well thickness. (a) Rashba spin-orbit coupling parameter, $\alpha$, for the InAs$_{1-x}$Sb$_{x}$ quantum well. (b,c) $g$-factor for the InAs$_{1-x}$Sb$_{x}$ quantum well. There is a non-monotonic behavior in both $\alpha$ and the $g$-factor going from pure InAs to pure InSb. By increasing the quantum well size, $\alpha$ decreases while the $g$-factor increases in magnitude.}
\label{fig:split-gfactor}
\end{center}
\end{figure*}

We use a standard 8-band ${\vec{k} \cdot \vec{p}}$ method~\cite{campos2019} to calculate the subband structure of the surface InAs$_{0.5}$Sb$_{0.5}$ QW. The quantum confinement along the growth direction was addressed by using the finite difference method with a discretization step of $0.5\,\textrm{nm}$ which is sufficient to achieve convergence. For computational efficiency, we neglect the impact of the  metal-semiconductor interface by modelling it as a hard-wall barrier acting as a confinement layer for the carriers. The material parameters were taken from Ref.~\citenum{vurgaftman2001band} while the bowing parameter for the InAs$_{0.5}$Sb$_{0.5}$ alloy was taken from Ref.~\citenum{webster2015measurement}.

Since the system has broken inversion symmetry, the energy dispersion, $\varepsilon_{n,\sigma}(k_z)$, is spin-split due to the Rashba SOC. In Figure~\ref{fig:split-gfactor}a we show the computed Rashba SOC parameter, $\alpha$, for the first conduction subband, computed as the linear slope of the energy difference $\Delta E = \varepsilon_{1,\sigma}(k_z) - \varepsilon_{1,\sigma^\prime}(k_z)$ very close to the $\Gamma$-point~\cite{campos2018spin}. In order to understand the quantum confinement as well as the effect of the alloy composition, $x$, we consider three InAs$_{1-x}$Sb$_{x}$ layer sizes and vary the composition $x$ from pure InAs to pure InSb. The gap at the InAs$_{1-x}$Sb$_{x}$/In$_{0.37}$Al$_{0.63}$Sb interface is a broken one, i.~e., the valence band edge is higher in energy than the conduction band edge and by increasing the InAs$_{1-x}$Sb$_{x}$ layer size the confined states' energies cross each other. In this situation, no spin-splitting was computed since the conduction and valence subbands crossed. Furthermore, the trend that the smaller the InAs$_{1-x}$Sb$_{x}$ layer size, the larger the Rashba parameter is due to the fact that the electron has a higher probability to be found near the interfaces than in the middle of the layer. Indeed, as we reduce the InAs$_{1-x}$Sb$_{x}$ layer size the Rashba SOC parameter becomes larger. We found that the highest value is around $\alpha = 0.35\,\textrm{eV/\AA}$ for the $10\,\textrm{nm}$ InAs$_{0.4}$Sb$_{0.6}$ layer, $\alpha = 0.2\,\textrm{eV/\AA}$ for the $20\,\textrm{nm}$ InAs$_{0.5}$Sb$_{0.5}$ layer, and $\alpha = 0.12\,\textrm{eV/\AA}$ for the $30\,\textrm{nm}$ InAs$_{0.6}$Sb$_{0.4}$ layer. Our calculation does not include electrostatic self-consistency, and since Fermi-level pinning, which one can expect to be composition dependent, would increase the asymmetry of the structure \cite{PhysRevB.63.155315}, our calculation provides a minimum for the spin-orbit coupling strength, and may not capture its full composition dependence.

The $g$-factor was computed using second order L\"owdin partitioning~\cite{winkler2003spin, tadjine2017universal}. In the bulk limit it converges to the Roth formula for an effective $g$-factor,
\begin{equation}
g^* = 2\left( 1- \frac{m_e}{m^*}\frac{\Delta_{SS}}{3\,E_g+2\,\Delta_{SS}}\right),
\label{eq:Roth}
\end{equation}
where $\Delta_{SS}$ is the spin-orbit splitting of the valence bands and $E_g$ is the energy gap, while 
$m_e$ and $m^{*}$ are the free and effective electron masses, respectively.
In Table~\ref{tb:gfactor}, we show the bulk $g$-factor for the In$_{0.37}$Al$_{0.63}$Sb barrier, InAs, InSb, and three selected InAs$_{1-x}$Sb$_{x}$ compositions. As we increase the composition, $x$, the band gap of the material decreases and since the main contribution to the $g$-factor comes from $1/E_g$~\cite{tadjine2017universal}, we obtain the largest $g$-factor values for compositions varying from $x=0.4$ to $x=0.6$. 

With quantum confinement, the $g$-factor is typically lower than the corresponding bulk value. This trend can also be inferred from Eq.~(\ref{eq:Roth}) since for a highly-confined system the effective band gap increases (as the energy difference from conduction to valence band also increases). We show the calculated $g$-factor for a confined system along the growth direction, $g_z$ in Figure~\ref{fig:split-gfactor}b, as well as perpendicular to the growth direction, $g_{x,y}$ in Figure~\ref{fig:split-gfactor}c. Due to the quantum confinement and SOC, the $g$-factor is anisotropic, i.e., $\Delta_g = g_{x,y} - g_z \neq 0$,~\cite{sandoval2016electron} with $g_z$ being larger in magnitude than $g_{x,y}$. Moreover, following the trend of the Roth formula in Eq.~(\ref{eq:Roth}), as we increase the size of the InAs$_{1-x}$Sb$_{x}$ layer, the $g$-factor also increases. We found that the largest $g$-factor occurs for a $30\,\textrm{nm}$ InAs$_{0.4}$Sb$_{0.6}$ QW and exceeds previous experimental results for a \SI{30}{nm} InSb QW which found in-plane $|g_{x,y}| = 26$ and out-of-plane $|g_z| = 52$ \cite{qu2016}. The above calculations show that there is a sweet spot in terms of QW width where $g$-factor and SOC are both strong. Motivated by this fact we focus the rest of our studies on 20~nm QWs.

\begin{table}
\caption{Bulk $g$-factor using Roth formula.}
\label{tb:gfactor}
\begin{tabular}{cc}

\hline 
Compound & $g^{*}$\tabularnewline
\hline 
\hline 
In$_{0.37}$Al$_{0.63}$Sb & -4.65\tabularnewline
InAs & -14.61\tabularnewline
InAs$_{0.6}$Sb$_{0.4}$ & -70.86\tabularnewline
InAs$_{0.5}$Sb$_{0.5}$ & -99.08\tabularnewline
InAs$_{0.4}$Sb$_{0.6}$ & -116.82\tabularnewline
InSb & -49.23\tabularnewline
\hline 
\end{tabular}
\end{table}

\begin{figure}[ht!]
    \centering
    \includegraphics[width=0.5\textwidth]{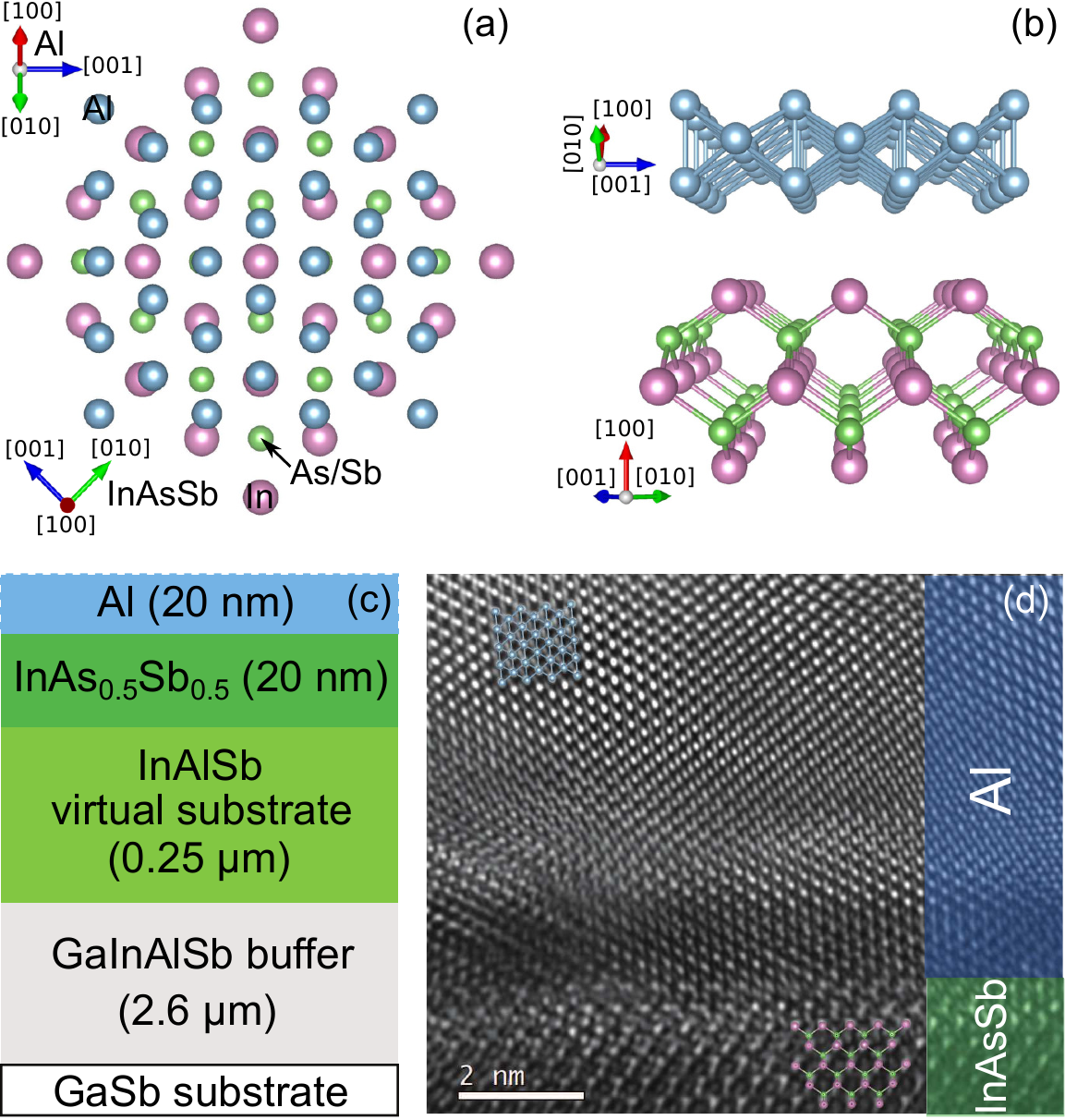}
    \caption{\label{fig:stack-TEM}(a) Unstrained Al on InAs$_{0.5}$Sb$_{0.5}$ with respective lattice constants \SI{4.05}{\angstrom} and \SI{6.27}{\angstrom}, projected onto the plane of growth. (b) Three-dimensional rendering of the Al-InAs$_{0.5}$Sb$_{0.5}$ interface from the perspective of the transmission electron microscope image below. (c) Layer diagram of the InAs$_{0.5}$Sb$_{0.5}$ surface quantum well with Al contact. (d) Cross-sectional transmission electron microscope image of the Al-InAs$_{0.5}$Sb$_{0.5}$ interface along the $\langle110\rangle$ zone axis with unstrained Al and InAs$_{0.5}$Sb$_{0.5}$ lattices overlaid.\cite{vesta}}
\end{figure}

Molecular beam epitaxy (MBE) growth of large-area InAs$_{0.5}$Sb$_{0.5}$ surface QWs in epitaxial contact to aluminum films can form the basis for combining proximity effect with high $g$-factor, strong SOC systems. Growth of semiconductor InAs$_{0.5}$Sb$_{0.5}$ is rather difficult since there is no insulating lattice-matched substrate immediately available. In this work, we pursue the process of compositional grading which allows growth of bulk unstrained, unrelaxed InAs$_{0.5}$Sb$_{0.5}$ of any composition onto GaSb, as previously reported \cite{Svensson12, Svensson11}. Following earlier work, our samples have a \SI{2.6}{\micro m} GaInAlSb compositional grade followed by a \SI{0.25}{\micro m} In$_{0.37}$Al$_{0.63}$Sb virtual substrate (VS) and a \SI{200}{\angstrom} InAs$_{0.5}$Sb$_{0.5}$ layer. Figure~\ref{fig:stack-TEM}c shows a schematic of the layers, while Figure~\ref{fig:stack-TEM}a shows a top view of the unstrained face-centered cubic Al lattice superimposed on the unstrained zincblende InAs$_{0.5}$Sb$_{0.5}$ lattice.  We grew the sample by solid-source molecular beam epitaxy in a modular Gen II system with the As and Sb delivered by valved cracker sources. Except during the Al layer deposition, the substrate temperature was measured with a K-space BandiT system operating in pyrometry mode.  Measurements of (004) triple-axis x-ray diffraction allowed us to verify the composition of the VS. We cannot examine the InAs$_{0.5}$Sb$_{0.5}$ layer, since it is too thin relative to the VS and the compositional grade, but test structures with thicker InAs$_{0.5}$Sb$_{0.5}$ layers were grown with this recipe and their composition was verified by X-ray diffraction.

\begin{figure}[ht]
\centering
\includegraphics[width=0.5\textwidth]{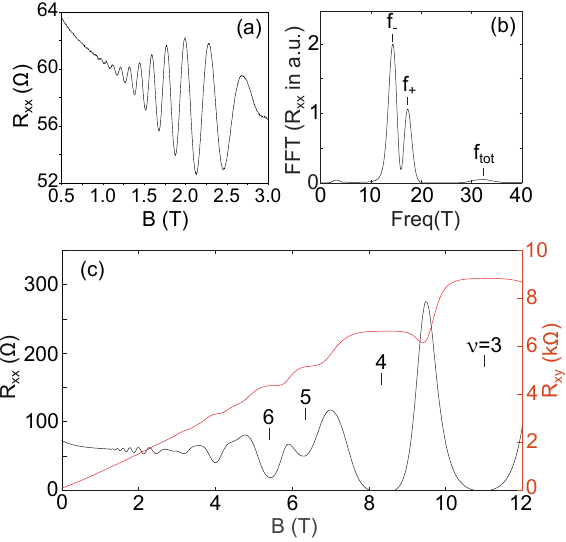}
\caption{\label{fig:transport}Transport measurements of a near-surface InAs$_{0.5}$Sb$_{0.5}$ quantum well with $n = \SI{8e11}{cm^{-2}}$ and $\mu = \SI{25e3}{cm^2/Vs}$ at \SI{1.5}{K}. (a) Beating of the longitudinal resistance at low magnetic field (a selected region of plot (c)). (b) Fast Fourier transform of the longitudinal resistance in the range \SI{1}{T}--\SI{5}{T}.  (c) Longitudinal (black) and Hall (red) resistance shown as a function of magnetic field.}
\end{figure}

For samples with Al, after the top InAs$_{0.5}$Sb$_{0.5}$ layer was grown, all shutters were closed and the sources were cooled to idling temperatures (around 300-400 $^\circ$C). The residual gases were pumped overnight, allowing the background pressure in the chamber to reach the \SI{e-11}{Torr} range. The next day, the sample was pointed towards the cryo-shroud for two hours and 40 minutes, allowing it to fall below \SI{0}{\celsius}. We deposited a \SI{200}{\angstrom} layer of Al onto the InAs$_{0.5}$Sb$_{0.5}$ surface at a growth rate of \SI{0.09}{\angstrom/s}. In this work, we present data from nominally identical structures, one with and one without an in-situ Al layer \cite{Sarney18}.
Figure~\ref{fig:stack-TEM}d shows a cross sectional transmission electron microscope image of the
interface between the InAs$_{0.5}$Sb$_{0.5}$ and Al layers along the $\langle110\rangle$ zone axis, while Figure~\ref{fig:stack-TEM}b shows a 3D rendering of the interface from the same perspective. The substrate and
InAs$_{0.5}$Sb$_{0.5}$ are oriented along a $\langle001\rangle$ growth direction. The Al film consists of large domains
predominately aligned along $\langle110\rangle$, tilted $\sim$4 degrees from the interfacial plane. The high
resolution images of this region and numerous others show that the d-spacing of the
growth direction planes is \SI{2.9}{\angstrom}, corresponding to that for Al along $\langle110\rangle$. The orientation
relationships of the crystal planes and the FFT pattern corresponds to Al examined at a
zone axis with a $\langle110\rangle$ growth direction. 

We studied the magnetoresistance of the InAs$_{0.5}$Sb$_{0.5}$ surface 2DEG without Al in a van der Pauw geometry.  Magnetotransport measurements were performed at $T=\SI{1.5}{K}$ using standard lock-in techniques and ac excitations $I_{ac}=\SI{50}{\nano A}$--$\SI{1}{\micro A}$ at frequencies below 100 Hz. We find mobilities of $\mu = 25,000$ cm$^2$/Vs at a carrier density $n = \SI{8e11}{cm^{-2}}$.

In the presence of strong SOC the Shubnikov-de Haas oscillations show two frequencies, signaling two Fermi surfaces, as can be seen in Figure~\ref{fig:transport}a, which suggests occupation of two spin-subbands. Figure~\ref{fig:transport}b shows the Fourier transform of these oscillations over the \SIrange{1}{5}{T} range. There are three clear peaks, which indicate spin-split subbands with frequencies $f_{+}$ = 17.2~T, $f_{-}$ = 14.2~T and a peak for the total frequency at $f_{tot}$ = \SI{33}{T}.  The densities can be directly calculated from $n_{\pm} = qf_\pm/h$ where $q$ is the electron charge and $h$ is Planck's constant. We obtain $n_{+} = \SI{4.2e11}{cm^{-2}}$ and $n_{-} = \SI{3.4e11}{cm^{-2}}$ with $n_{tot} = \SI{7.6e11}{cm^{-2}}$ which agrees with the Hall data shown in  Figure~\ref{fig:transport}c. This suggests the spin-split subband separation is very large as expected for InAs$_{0.5}$Sb$_{0.5}$. If this splitting were all due to the linear Rashba SOC term, we would obtain its parameter as $\alpha = (\Delta n \hbar^{2}/m^{*})\sqrt{\pi/[2(n_{tot}-\Delta n)]} = \SI{0.8}{eV/\angstrom}$, where $\Delta n = |n_+ - n_-|$, assuming a band mass of $m^{*} = 0.011 \,m_e$ at 50\% composition \cite{massInAsSb}. Our ${\vec{k} \cdot \vec{p}}$ calculation for this QW width predicts $\alpha=\SI{0.2}{eV/\angstrom}$ which is lower than the $\alpha=\SI{0.8}{eV/\angstrom}$ estimated from extracted parameters, suggesting there are  contributions from Dresselhaus SOC terms in Sb compounds \cite{Santos2008}. We also note that a Schr\"odinger-Poisson calculation for our 20~nm QW showed one electronic subband is occupied.

\begin{figure}[htp]
\centering
\includegraphics[width=0.5\textwidth]{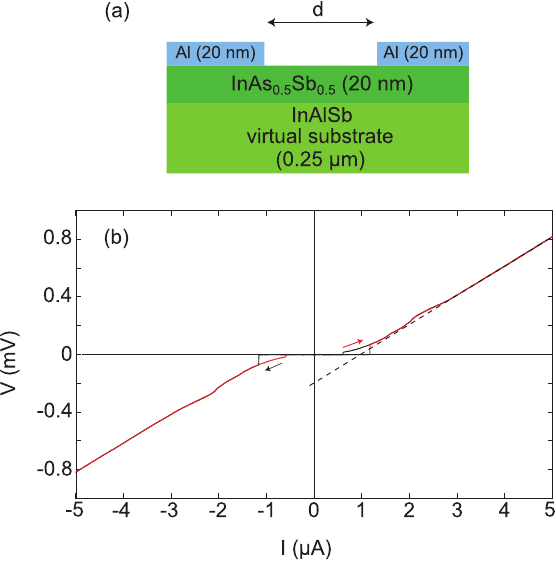}
\caption{\label{fig:JJ-IV}(a) Schematic of the Josephson junctions on InAs$_{0.5}$Sb$_{0.5}$ with separation $d$ between Al contacts. (b) Current-voltage characteristics of the $d=\SI{500}{\nano\meter}$ junction at \SI{20}{mK}. The dashed line is an extrapolated linear fit of the region $eV > 2\Delta_{Al}$.}
\end{figure}

We further characterize the superconducting proximity effect in a Josephson junction (JJ) on an InAs$_{0.5}$Sb$_{0.5}$ 2DEG with epitaxial Al contacts, as depicted schematically in Figure~\ref{fig:JJ-IV}a. We fabricated Josephson junctions via electron beam lithography followed by selective wet etching (Transine type D) to remove a thin strip of Al. The junction is \SI{4}{\micro m} wide and has a \SI{500}{nm} length separation between the superconducting electrodes. Measurements were performed in a dilution fridge with a mixing chamber temperature of \SI{7}{mK} and an estimated electron temperature of \SI{20}{mK}. We used a 4-point measurement geometry and standard dc current bias techniques.  The I-V characteristic of the junction is shown in Figure~\ref{fig:JJ-IV}b. The voltage drop across the junction is zero (the supercurrent) up to a critical value of driving current denoted the critical current, $I_{c} = \SI{1.16}{\micro A}$.

The quality of the device can be characterized by a study of the $I_cR_N$ and $I_{ex}R_N$ products, where $R_N$ is the normal resistance of the JJ. The excess current $I_{ex}$ is the difference between the measured current through the junction and the expected current based on the junction's $R_N$. This occurs due to Andreev reflections and depends primarily on interface transparency. The critical current $I_{c}$ is the amount of current that can be carried by Andreev bound states through the junction with zero resistance. $I_{c}$ requires coherent charge transport across the semiconductor region and is therefore a measure of both interface transparency and 2DEG mobility.

The junction is neither clearly ballistic ($l_e \gg d$) nor diffusive ($d \gg l_e$) since the mean free path $l_{e} \sim$ $370~$ nm is of the order of the contact separation. The mean free path is obtained from the transport measurements in a van der Pauw geometry presented earlier. Using the diffusive expression for the Thouless energy $E_T = \frac{\hbar\,D}{d^2}$ with D the diffusion constant, we find $E_T \sim 1.1 meV$. This value, which is likely underestimated since our junction is not deeply diffusive, is about 5 times larger than the superconducting gap $\Delta_{Al}=210$~$\mu$eV which we extract from the BCS relation $\Delta_{Al}= 1.75k_BT_c$ (see Supporting Information). As a consequence our junction is close to the short limit.

The Andreev process that carries the supercurrent across the Sm region is characterized by the induced gap $\Delta_{\rm ind}$ in the Sm below the S, rather than the bulk Al gap, $\Delta_{Al}$. To characterize an S-Sm-S junction in the short limit, the product of the critical current and the normal state resistance, which is related to the gap via $I_{c}R_N = \eta \Delta_{Al}/e$, is often used, where $\eta$ is a constant of order unity. Experimentally, we find $I_{c}R_N = \SI{270}{\micro V}$ where $I_c = \SI{1.16}{\micro A}$, $R_N = \SI{230}{\ohm}$, in the junction with $d=\SI{500}{nm}$ contact separation at T = 20~mK, consistent with previous results in InAsSb nanowires \cite{Sestoft18}.  The product of I$_{c}$R$_N$ can be compared to theoretical values for fully transparent junctions in the short ballistic and short diffusive limits, for which $\eta$ is $\pi$ and $1.32(\pi/2)$, respectively \cite{IcRnreview}. For our sample, we find $I_{c}R_{N} $ is $37\%$ of the ballistic limit and $57\%$ of the diffusive limit. This results are comparable with what has been observed in InAs 2DEG for similar contact separations \cite{Mayer19Proximity}.

Due to the high mobility of the InAs$_{0.5}$Sb$_{0.5}$ channel he supercurrent persists at longer separations. At $d=\SI{1}{\micro\meter}$ separation we still observe a substantial supercurrent $I_c = \SI{570}{nA}$ with $I_{c}R_N = \SI{280}{\micro V}$; raw data is presented in the supplementary information.

High interface transparency corresponds to a high probability of Andreev reflection at the interface. Since the Sm extends under the S regions, the interface between Sm and S should be highly transparent due to the large area of contact and in-situ epitaxial Al growth \cite{MortenNatureComm16}. The Andreev process that carries supercurrent across the Sm region is characterized by the excess current $I_{ex}= I-V/R_N$ through the junction. Excess current does not require coherent charge transport across the junction as it follows simply from charge conservation at the S-Sm interfaces. $I_{ex}$ can be calculated by extrapolating from the high current normal regime to zero voltage as shown in Figure~4b with a dotted line. The excess current in our sample is found to be $I_{ex}=1\mu$A.

When considering interface quality the more relevant quantity is the product $I_{ex}R_N$. The product $I_{ex}R_N$ can be compared to the superconducting gap with the relation $I_{ex}R_N=\eta'\Delta_{Al}/e$. In the case of a fully transparent S-Sm interface $\eta'=1.467$ for a diffusive junction \cite{IexeRnpaper} and $\eta'=8/3$ for a ballistic junction. For our sample, $I_{ex}R_N=230$~$\mu V$, which is close to values reported in InAs 2DEG \cite{Mayer19Proximity}. This value is $35\%$ of the ballistic limit and $65\%$ of the diffusive value for our 500~nm JJ. The Octavio–Tinkham–Blonder–Klapwijk theory allows to link the ratio $\frac{I_{ex}R_N}{\Delta}$ to the interface transparency. Using equation (25) of Ref \citenum{niebler_analytical_2009}, we can extract the effective scattering parameter $Z = 0.58$, leading to a 75\% probability of Andreev reflection at zero energy. This value is similar to transparencies observed in InAs nanowires \cite{gharavi_nbinas_2017}.


In conclusion, we have demonstrated a robust superconducting proximity effect in two-dimensional epitaxial Al-InAs$_{0.5}$Sb$_{0.5}$ systems. Using an optimized MBE growth we have achieved both high electron mobilities in InAs$_{0.5}$Sb$_{0.5}$ and successful epitaxial growth of thin film Al. Outstanding transport properties were confirmed in the normal and superconducting state by Shubnikov-de Haas oscillations and current-voltage measurements, which establish strong spin-orbit coupling and large critical current in Josephson junctions. Remarkably, the latter property, made possible by high interface transparency, is consistent with a large proximity-induced superconducting gap of $\sim 270$ $\mu$eV in InAs$_{0.5}$Sb$_{0.5}$. The supercurrent between two Al contacts can be sustained in InAs$_{0.5}$Sb$_{0.5}$ across at least 1000 nm.

While these results clearly indicate that InAs$_{0.5}$Sb$_{0.5}$-based junctions provide a suitable platform in which to explore topological superconductivity, they also have broader implications. We expect that spin-orbit coupling in InAs$_{0.5}$Sb$_{0.5}$ could be further controlled through electrostatic gating or magnetic structures to modify quantum transport both in the normal and superconducting state.

\begin{suppinfo}
Temperature dependence of supercurrent and $I_cR_N$ products; Characterization of supercurrent in \SI{1}{\micro\meter} Josephson junction.
\end{suppinfo}

\section{Acknowledgement}
This work was partially supported by NSF DMR 1836687, the US Army research office, US ONR N000141712793, NSF
ECCS-1810266, the University at Buffalo Center for Computational Research, and the ARO/LPS Quantum Computing Graduate Research Fellowship (QuaCGR BAA W911NF-17-S-0002).

\section{Authors contributions} W.M, M.H. performed the measurements and analysis, J.Y. and M.C.D. helped with fabrication of the devices with J.S. providing input. K.S.W. and J.S. designed the stack. W.L.S and S.P.S grew the epitaxial Al/InAs heterostructures. A.C.L. performed TEM analysis. T.C. developed the simulation model and carried out the simulations. J.S. conceived the experiment. All authors contributed to interpreting the data. The manuscript was written by W.M, W.F.S., M.C.D., I.{\v{Z}}., and J.S. with suggestions from all the other authors.

\section{Notes}
The authors declare no competing financial interest. 

\bibliography{References_Shabani_Growth}

\end{document}